\pdfoutput=1
\documentclass[preprint,12pt]{elsarticle}

\usepackage{graphicx}
\usepackage{amssymb}
\usepackage[toc,page]{appendix}
\usepackage[T1]{fontenc}
\usepackage[swedish,english]{babel}
\usepackage{amsmath,amssymb,bbm}
\usepackage{amsthm}
\usepackage{amsfonts}
\usepackage{ae}
\usepackage{verbatim}
\usepackage{icomma}
\usepackage{dsfont}
\usepackage{bbm}
\usepackage{graphicx}
\usepackage{mathtools}

\usepackage{lineno}

\newcommand{\mikroeV}{\mu\text{eV}}
\newcommand{\meV}{\text{meV}}
\newcommand{\eV}{\text{eV}}
\newcommand{\keV}{\text{keV}}
\newcommand{\MeV}{\text{MeV}}

\newcommand{\K}{\text{K}}
\newcommand{\s}{\text{s}}
\newcommand{\cm}{\text{cm}}
\newcommand{\gram}{\text{g}}
\newcommand{\de}{\text{d}}
\newcommand{\kms}{\text{km/s}}

\journal{JCAP}

\begin{document}

\begin{frontmatter}


\title{21 cm cosmology and spin temperature reduction via spin-dependent dark matter interactions}



\author{Axel Widmark\footnote{Corresponding author, axel.widmark@fysik.su.se}}

\address{Stockholm University, Sweden}

\begin{abstract}
The \emph{EDGES} low-band experiment has measured an absorption feature in the cosmic microwave background radiation (CMB), corresponding to the 21 cm hyperfine transition of hydrogen at redshift $z \simeq 17$, before the era of cosmic reionization. The amplitude of this absorption is connected to the ratio of singlet and triplet hyperfine states in the hydrogen gas, which can be parametrized by a spin temperature. The \emph{EDGES} result suggests that the spin temperature is lower than the expected temperatures of both the CMB and the hydrogen gas. A variety of mechanisms have been proposed in order to explain this signal, for example by lowering the kinetic temperature of the hydrogen gas via dark matter interactions. We introduce an alternative mechanism, by which a sub-GeV dark matter particle with spin-dependent coupling to nucleons or electrons can cause hyperfine transitions and lower the spin temperature directly, with negligible reduction of the kinetic temperature of the hydrogen gas. We consider a model with an asymmetric dark matter fermion and a light pseudo-vector mediator. Significant reduction of the spin temperature by this simple model is excluded, most strongly by coupling constant bounds coming from stellar cooling. Perhaps an alternative dark sector model, subject to different sets of constraints, can lower the spin temperature by the same mechanism.
\end{abstract}

\begin{keyword}
21 cm cosmology \sep Dark matter


\end{keyword}

\end{frontmatter}


\section{Introduction}\label{sec:introduction}

The epoch between the formation of the cosmic microwave background radiation (CMB) and reionization is commonly referred to as the cosmic dark ages. In this epoch, matter is largely uniform and transparent to radiation. An exception to its transparency is the hyperfine transition of the hydrogen atom, between its electron ground state triplet and singlet states. The transition energy corresponds to a photon wavelength of 21 cm, which can interact with the CMB. Excitations and deexcitations are also caused by gas collisions and Ly$\alpha$ radiation. The relative abundance of singlet and triplet states can be parametrized by a spin temperature, which will tend to the temperature associated with the dominant process of hyperfine transitions. If the spin temperature is lower than the CMB temperature, this will cause an absorption feature in the CMB. For the first time, such a signal has been observed by the \emph{EDGES} experiment before the epoch of reionization \cite{Bowman:2018yin}. The results suggest that the spin temperature at redshift $z\simeq 17$ is lower than the expected temperatures of both the CMB and the hydrogen gas. While the validity of this measurement is still up for debate \cite{2018arXiv180501421H}, it will hopefully be independently tested in the near future. If the result persists, there will be a definite need to explain it with new, possibly dark sector, physics.\footnote{Non-dark sector explanations have also been proposed, such as an enhanced soft photon background \cite{Feng:2018rje,Ewall-Wice:2018bzf,Fraser:2018acy,Pospelov:2018kdh}, e.g. due
to decaying or annihilating particles \cite{Chluba:2015hma}.}

A number of articles have discussed the possibility that dark matter can cool the hydrogen gas, which in turn couples to the spin temperature. In a standard scenario, the hydrogen gas has cooled adiabatically since its thermal decoupling from the CMB. Because the dark matter is significantly colder, spin-independent interactions with a $v^{-4}$ velocity dependent cross section can cool the gas temperature around redshift $z\simeq 17$. Such models are constrained to a sub-dominant component of milli-charged dark matter, constituting only about one per cent of the total dark matter abundance \cite{Berlin:2018sjs,Munoz:2018pzp,Munoz:2018jwq,Barkana:2018cct}.

We propose an alternative mechanism by which spin-dependent dark matter interactions can lower the spin temperature, not by lowering the kinetic temperature of the hydrogen gas but by directly causing hyperfine transitions in the hydrogen atom. We consider a sub-GeV dark matter fermion with spin-dependent interactions with either nucleons or electrons, mediated by a light pseudo-vector. Due to the mass difference between hydrogen and the dark matter fermion, momentum transfer is strongly suppressed, such that the hydrogen gas temperature is only marginally affected by dark matter interactions. However, because interactions are spin-dependent, they can excite or deexcite the hyperfine triplet state. Because this is an inelastic collision, excitations become energetically impossible in the low velocity limit, while deexcitations are significantly enhanced.

The paper is outlined as follows. In Sec. \ref{sec:theory} we present the theoretical background for spin temperature thermodynamics, our dark sector particle model, and the relevant cross sections and scattering rates. We discuss constraints to the dark sector parameter space in Sec. \ref{sec:limits}. In Sec. \ref{sec:results}, we present the results in terms of how dark matter can affect the spin temperature. Finally, we discuss and conclude in Sec. \ref{sec:discussion}.

\section{Theory}\label{sec:theory}

In this section we present the theoretical background necessary to calculate the spin temperature under the influence of CMB radiation, gas collisions, Ly$\alpha$ radiation, and dark sector interactions. A review on the subject of 21 cm cosmology, from where most of the thermodynamic formalism in this article is taken, can be found in \cite{Pritchard:2011xb}.

\subsection{Spin temperature}\label{sec:spin_temperature}

The spin temperature, although not a true thermodynamic temperature, is defined by the relative abundance of triplet and singlet states in the hydrogen gas, according to
\begin{equation}\label{eq:Ts_definition}
	\frac{n_1}{n_0} = 3\exp\left(-\frac{T_\star}{T_s}\right),
\end{equation}
where $n_0$ and $n_1$ are the singlet and triplet number densities, and
\begin{equation}
    T_\star = 0.068~\K,
\end{equation}
is the temperature that corresponds to the hyperfine transition energy
\begin{equation}
    E_\star=5.9~\mikroeV.
\end{equation}

The spin temperature can be derived as follows. Because all relevant processes are much faster than the spontaneous deexcitation of the triplet state, we can assume steady state. The ratio of hydrogen triplet and singlet states is equal to the ratio of excitation and deexcitation probabilities ($P_{01}$ and $P_{10}$). The three main processes that cause these transitions are interactions with the CMB around the 21 cm line, gas collisions, and Ly$\alpha$ radiation produced by the first stars, denoted by indices $\gamma$, $k$, and $\alpha$. The excitation and deexcitation probabilities are related to each other via simple functions of the temperature, giving \cite{1958PIRE...46..240F}
\begin{equation}\label{eq:Ts}
\begin{split}
    & 3\exp\left(-\frac{T_\star}{T_s}\right) = \frac{n_1}{n_0} =
    \frac{P_{01}^\gamma+P_{01}^K+P_{01}^\alpha}{P_{10}^\gamma+P_{10}^K+P_{10}^\alpha} =
    \\
    & = \dfrac{3 A_{10} \dfrac{T_\gamma}{T_\star} +
    3\exp \left( -\dfrac{T_\star}{T_g} \right) P_{10}^K+3\exp \left( -\dfrac{T_\star}{T_g} \right)P_{10}^\alpha}
    {A_{10} \left(1+\dfrac{T_\gamma}{T_\star}\right) + P_{10}^K+P_{10}^\alpha},
\end{split}
\end{equation}
where $A_{10}=2.85\times 10^{-15}~\s^{-1}$ is the spontaneous decay rate of the triplet state, and $T_{i=\{\gamma,K\}}$ are the temperatures of the CMB and hydrogen gas (the color temperature of the Ly$\alpha$ radiation is closely related to $T_K$ by scattering recoil).\footnote{Although not adopted here, it is common to rewrite Eq.~\eqref{eq:Ts} using the simplification
\begin{equation}
	\exp\left(-\frac{T_\star}{T}\right) \simeq \left(1-\frac{T_\star}{T}\right),
\end{equation}
which is valid for $T \gg T_\star$. This gives
\begin{equation}\label{eq:thermo_no_DM}
	T_s^{-1} = \frac{T_\gamma^{-1}+(x_K+x_\alpha) T_K^{-1}}{1+x_K+x_\alpha},
\end{equation}
where $x_{i=\{K,\alpha\}}$ are coupling strengths equal to
\begin{equation}
    x_i = \frac{T_\star}{A_{10} T_K} P^i_{10}.
\end{equation}
}

The transition probability of going from the triplet to the singlet state due to gas collisions is equal to
\begin{equation}
	P^K_{10} = n_H \kappa_{10}^H+n_e \kappa_{10}^e,
\end{equation}
where $n_H$ and $n_e$ are the number densities of hydrogen atoms and free electrons, and $\kappa_{10}^H$ and $\kappa_{10}^e$ are their respective scattering rates \cite{1969ApJ...158..423A,0004-637X-622-2-1356,2007MNRAS.374..547F,doi:10.1111/j.1365-2966.2007.11921.x}. Scattering by free protons can be neglected as it is always sub-dominant to free electrons. Approximate functions for $\kappa_{10}^H$ and $\kappa_{10}^e$, valid in our regime of temperatures, are \cite{Kuhlen:2005cm,Liszt:2001kh}
\begin{equation}
\begin{split}
	\kappa_{10}^H & = 3.1\times 10^{-11} T_K^{0.357}\exp\Bigg( -\frac{32}{T_K} \Bigg)~\frac{\cm^3}{\s}, \\
    \kappa_{10}^e & = 10^\wedge \Bigg\{ -9.607+0.5\log_{10} (T_K)\exp\Bigg[-\frac{\log_{10} (T_K)^{9/2}}{1800} \Bigg] \Bigg\}~\frac{\cm^3}{\s}.
\end{split}
\end{equation}
The number density of hydrogen is
\begin{equation}
f_H \frac{\Sigma_b \rho_c}{m_H},
\end{equation}
where $f_H$ is the mass fraction of hydrogen with respect to the total baryonic density, $\Sigma_b$ and $\rho_c$ are the baryon abundance and critical density of the $\Lambda$CDM model, and $m_H$ is the hydrogen mass. The fraction of free electrons is calculated with \texttt{recfast} \cite{Wong:2007ym}.

Before the era of cosmic reionization, around redshift $z\simeq 15$--$20$, the first stars of the universe starts heating the hydrogen gas and also couples the spin temperature to the gas temperature via Ly$\alpha$ radiation, through the Wouthuysen-Field effect \cite{1952AJ.....57R..31W}. The theoretical predictions for when this heating begins or how the coupling strength evolves with time are not very well constrained \cite{2017MNRAS.472.1915C}. We do not specify any model for the heating of the hydrogen gas and Ly$\alpha$ coupling of the spin temperature, but suffice by saying that it comes into significant effect around a redshift of $z \simeq 17$. For dark matter models with spin-independent interactions, lowering the spin temperature is contingent on a significant amount of Ly$\alpha$ radiation being produced, coupling the spin and hydrogen gas temperatures, before any significant heating of the hydrogen gas takes place. The dark sector model considered in this work is qualitatively different, in that it lowers the spin temperature directly. For this reason, the specific behavior of star formation and Ly$\alpha$ radiation is not detrimental to the results of this work.

\begin{figure}[tbp]
\centering
\includegraphics[width=1.\textwidth,origin=c]{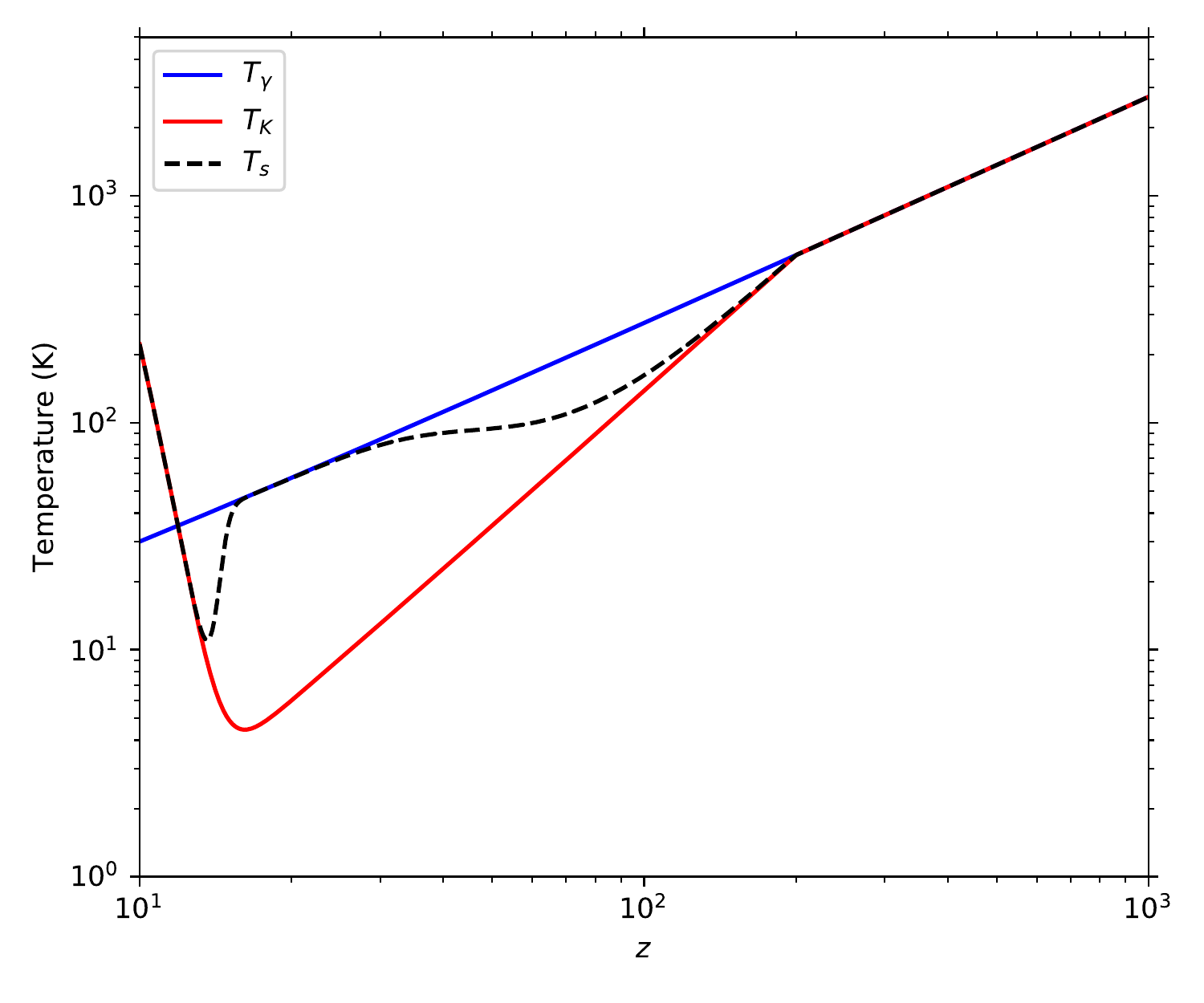}
\caption{\label{fig:Tspin}CMB temperature ($T_\gamma$), kinetic hydrogen gas temperature ($T_K$), and spin temperature ($T_s$), as a function of redshift ($z$). No dark sector is influencing the spin temperature in this figure.}
\end{figure}

The evolution of the spin temperature as a function of redshift is visible in Fig.~\ref{fig:Tspin}, for the case where no dark matter is affecting the spin gas. At high redshifts, the CMB and gas temperatures are thermally coupled through a small fraction of free electrons, following $T_{\gamma,K} \propto 1+z$. At redshift $z \simeq 200$, the gas temperature decouples from the CMB and cools adiabatically at a rate $T_K \propto (1+z)^2$. For some time, gas collisions are the dominant process of hyperfine transitions and the spin temperature follows the gas temperature $T_K$. As the gas temperature drops and collisions become less frequent, the spin temperature couples to the CMB. At redshift $z\simeq 17$, star formation heats the hydrogen gas and Ly$\alpha$ radiation couples the spin and gas temperatures.

In this work, we introduce a dark matter species that causes hyperfine transitions via spin-dependent interactions. The above equations have been standard for 21 cm cosmology, but now we modify Eq.~\eqref{eq:Ts}. We add the dark matter induced excitation and deexcitation probabilities in the nominator and denominator of Eq.~\eqref{eq:Ts}, giving
\begin{equation}
    3\exp\left( -\frac{T_\star}{T_s} \right) =
    \dfrac{3\dfrac{A_{10} T_\gamma}{T_\star} +
    3\exp\left( -\dfrac{T_\star}{T_g} \right) (P_{10}^K+P_{10}^\alpha)
    +P_{01}^\chi}
    {A_{10} \left(1+\dfrac{T_\gamma}{T_\star}\right) + P_{10}^K+P_{10}^\alpha+P_{10}^\chi}.
\end{equation}
The deexcitation rate due to dark matter interactions is given by
\begin{equation}\label{eq:deexcitation_rate}
	P_{10}^\chi = n_\chi \int_{0}^\infty \sigma^+_{\chi N}(v)\; v f_{\tilde{T}}(v)\; \de v,
\end{equation}
where $n_\chi$ is the number density of dark matter particles, $\sigma^+_{\chi N}$ is the inelastic deexcitation cross section of Eq.~\eqref{eq:cross_section_full}, and $f_{\tilde{T}}(v)$ is the Maxwell-Boltzmann velocity distribution with a velocity dispersion parametrized by an effective temperature $\tilde{T}$ (see section \ref{sec:effective_temperature}). The excitation rate $P_{01}^\chi$ is equivalently expressed,
\begin{equation}\label{eq:excitation_rate}
	P_{01}^\chi = 3 n_\chi \int_{\sqrt{2 E_\star / \mu}}^\infty \sigma^-_{\chi N}(v)\; v f_{\tilde{T}}(v)\; \de v,
\end{equation}
where $\sigma^-_{\chi N}$ is the inelastic excitation cross section, and the factor 3 comes from the multiplicity of the triplet state. The lower bound of the integral is different from the deexcitation case, because excitation through lower collisional velocities is energetically impossible. The excitation and deexcitation cross sections are presented in Sec.~\ref{sec:cross_section}. The excitation and deexcitation rates, $P_{01}^\chi$ and $P_{10}^\chi$, are related via a temperature function, analogous to the relation between the rates due to gas collisions.

\subsection{Dark sector particle model}\label{sec:particle_model}

We consider a dark sector particle model consisting of a fermion $\chi$, with a mass between $1~\keV$ and $10~\MeV$, with spin-dependent coupling to nucleons or electrons through a pseudo-vector mediator $V$, with a mass between $1~\meV$ and $10~\eV$. The interaction terms of the Lagrangian are written
\begin{equation}
	\mathcal{L} \supset
    g_\chi V_\mu \bar{\chi}\gamma^\mu \gamma^5 \chi +
    g_N V_\mu \bar{N}\gamma^\mu \gamma^5 N,
\end{equation}
where $N$ is a baryonic nucleon, and $g_\chi$ and $g_N$ are coupling constants. For leptophilic dark matter, the nucleon $N$ is replaced with an electron $e$.

The mediator $V$ is significantly lighter than the dark matter fermion $\chi$. In order to prevent pair annihilations that would deplete the universe of the dark matter fermion component, this fermion must be asymmetric \cite{Petraki:2013wwa,Zurek:2013wia}.

The cross section of elastic nucleon-dark matter interactions at non-relativistic velocities is equal to
\begin{equation}\label{eq:cross_section}
	\sigma_{\chi N}(v) = \frac{g_\chi^2 g_N^2 m_\chi^2}{4\pi [ (m_\chi v/c)^2 + m_V^2 ]^2}.
\end{equation}
where $v$ is the collisional velocity and $c$ is the speed of light. This cross section has a $v^{-4}$ velocity dependence, down to $v/c \gtrsim m_V/m_\chi$. This lower bound corresponds to when the de Broglie wavelength of the incoming dark matter particle is longer than the range of the force, set by the mass of the force carrying particle.

\subsubsection{Inelastic scattering cross section}\label{sec:cross_section}

In a collision between particles where one of the particles is excited or deexcited, there will be a deficit or surplus of kinetic energy in the outgoing particle trajectories. If the collisional energy is sufficiently low, excitations become kinetically forbidden and deexcitations become amplified. This can be described by a form factor, which is an additional contribution to the elastic scattering cross section.

Given a surplus of outgoing kinetic energy $\delta E$ and assuming non-relativistic velocities, the form factor is given by the ratio of in-going and out-going phase space volumes. Written in terms of collisional velocity $v$, the form factor for hydrogen-dark matter inelastic scattering is equal to
\begin{equation}
	F(v,\delta E) = \Theta \left(\frac{\mu v^2}{2}  + \delta E \right)\frac{\mu v^2/2 + \delta E}{\mu v^2/2},
\end{equation}
where $\mu \equiv m_H m_\chi /(m_H + m_\chi)$ is the reduced mass of the hydrogen atom and dark matter fermion, and $\Theta(\mu v^2/2+\delta E)$ is the Heaviside step function, only relevant in the excitation case for which $\delta E<0$. In the limit $m_\chi \ll m_H$, we have that $\mu = m_\chi$, and almost all of the surplus energy $\delta E$ is carried away by the dark matter particle.

\begin{figure}[tbp]
\centering
\includegraphics[width=.833\textwidth,origin=c]{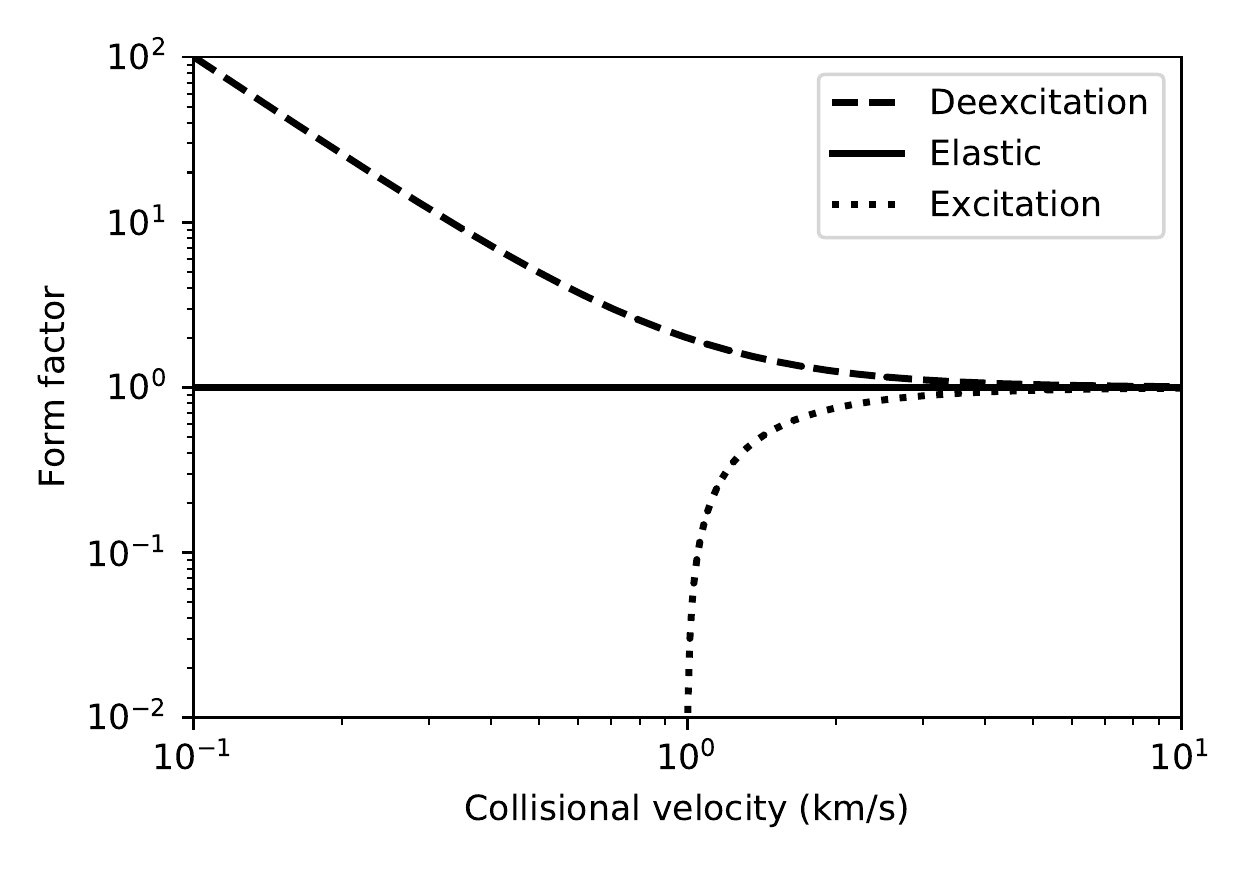}
\caption{\label{fig:form_factor}Form factor for excitation ($F_{01}$) and deexcitation ($F_{10}$) of the hyperfine transition as a function of collisional velocity, for interactions between hydrogen and a dark matter particle of mass $m_\chi = 1~\MeV$. Also shown is the form factor of elastic collisions, which is constant.}
\end{figure}

Excitations are quenched and deexcitations are significantly amplified when the in-going collisional energy is smaller than the hyperfine transition energy $E_\star$. This is illustrated in Fig.~\ref{fig:form_factor}, where the form factor is shown as a function of collisional velocity.

\begin{figure}[tbp]
\centering
\includegraphics[width=.833\textwidth,origin=c]{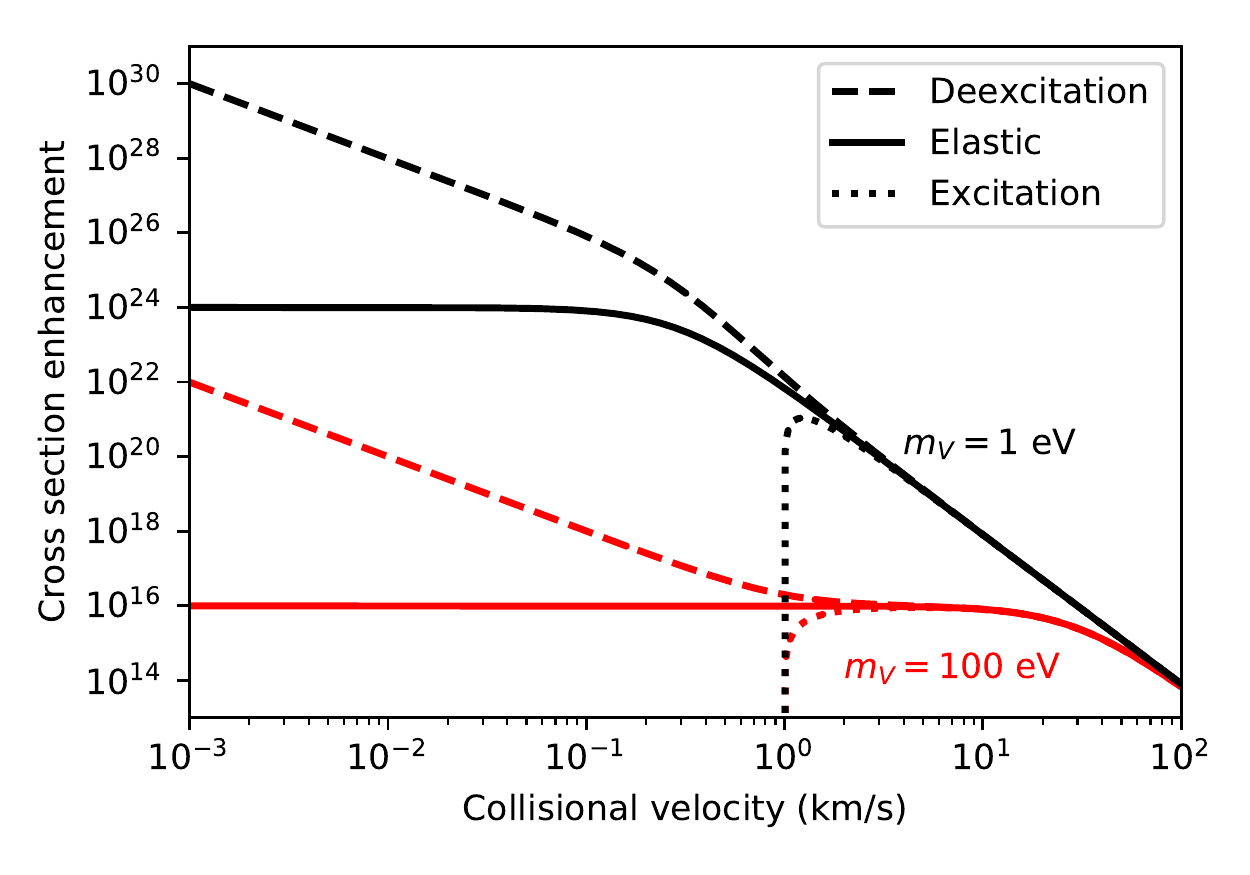}
\caption{\label{fig:cross_section_enhancement}Cross section enhancement, relative to the cross section of relativistic scattering, with both $v^{-4}$ and inelastic scattering amplifications. The dark matter mass is set to $m_\chi=1~\MeV$, and the mediator mass is $m_V=1~\eV$ (black) or $m_V=100~\eV$ (red). Significant amplification due to inelastic scattering starts around $v=1~\kms$. The knee of $v^{-4}$ enhancement is located at $v=(m_V/m_\chi) c$.}
\end{figure}

There is a negligible contribution to the cross section coming from the virtual boson propagator now carrying additional momentum in deexcitation, and reduced momentum in the excitation case. This can safely be ignored, as $m_V \gg E_\star$. The full cross section of inelastic scattering is
\begin{equation}\label{eq:cross_section_full}
	\sigma^\pm_{\chi N} = F(v,\pm E_\star)\,\sigma_{\chi N}(v),
\end{equation}
where $\sigma_{\chi N}(v)$ is the elastic cross section of Eq.~\eqref{eq:cross_section}. The cross section enhancement as a function of collisional velocity, with respect to $v=c$, is shown in Fig.~\ref{fig:cross_section_enhancement}.

There is no significant Sommerfeld enhancement of the scattering cross section. Slatyer \cite{2010JCAP...02..028S} has calculated the annihilation cross section for dark matter particles with inelastic interactions mediated by a scalar $\phi$, with coupling constant $g$ and energy splitting $\delta E$. They find that Sommerfeld enhancement can only be significant if the following conditions are fulfilled:
\begin{equation}\label{eq:sommerfeld_condition}
\begin{rcases}
    v/c \\
    \sqrt{\delta E / m_\chi} \\
    m_\phi / m_\chi
\end{rcases} \lesssim g^2.
\end{equation}
The hydrogen-dark matter interactions considered in this work have a similar behavior, although we replace $g^2$ by $g_N g_\chi$. Given the coupling constant limits that are discussed in Sec.~\ref{sec:limits}, these conditions are not fulfilled; for example, the second condition is clearly broken, as $\sqrt{E_\star / m_H} \gg 10^{-9} \gg g_N g_\chi$, where we have used the hydrogen mass which is the heavier particle in the interaction.

\subsection{Effective temperature of hydrogen-dark matter collisions}\label{sec:effective_temperature}

Given a hydrogen gas and a dark matter gas, following Maxwell-Boltzmann velocity distributions with velocity dispersions $\langle v_H ^2 \rangle$ and $\langle v_\chi ^2 \rangle$, the relative velocities between a hydrogen and dark matter has dispersion $\langle \tilde{v}^2 \rangle = \langle v_H ^2 \rangle + \langle v_\chi ^2 \rangle$, also following a Maxwell-Boltzmann distribution. The two gases are not in equilibrium with each other and have separate temperatures. Yet, the average collisional velocity between hydrogen and dark matter can be parametrized by an effective temperature, given by the equation
\begin{equation}
	\tilde{T}(m_H^{-1}+m_\chi^{-1}) = T_K m_H^{-1} + T_\chi m_\chi^{-1}.
\end{equation}
In the limit that $m_\chi \ll m_H$ and $T_\chi/T_H \ll m_\chi/m_H$, we get that
\begin{equation}
	\tilde{T} = \frac{m_\chi}{m_H}T_K,
\end{equation}
which can be significantly lower than the spin temperature.

In an ideal case, the dark matter gas would be absolutely inert, although this is not entirely realistic. There can be relative bulk motions of the hydrogen and dark matter gases, with velocities comparable to the sound speed of the hydrogen gas \cite{2010PhRvD..82h3520T}. Furthermore, even if we assume the dark matter to be infinitely cold at high redshift ($z \simeq 1100$), it can heat up due to interactions with the hydrogen gas. The energy transfer from the hydrogen is kinematically disfavored due to the relative mass difference between hydrogen and the dark matter fermion; for this reason, heating is dominated by inelastic collisions where energy is transferred from the spin gas. The energy contained in the spin gas is very small, but the effective temperature $\tilde{T}$ is sensitive even to a very small increase in the dark matter temperature $T_\chi$. It will be evident in Sec.~\ref{sec:results} that heating of the dark matter gas is more significant than the relative bulk motion of hydrogen, such that the latter is negligible.

The net heating of the dark matter temperature, at some redshift, is
\begin{equation}
    T_\chi(z) =
    \frac{2}{3 k_B} (1+z)^2
    \int_{z}^{1100} \frac{\de E_\chi}{\de t}
    \frac{\de t}{\de z'} (1+z')^{-2} \de z'.
\end{equation}
The redshift factors $(1+z)^2$ and $(1+z')^{-2}$, outside and inside the integral, are due to adiabatic cooling by expansion. The assumption that the dark matter is inert at the time of recombination sets the upper bound of the integral. As we shall see in Sec.~\ref{sec:results}, heating is negligible for $z\gtrsim100$. Because the relevant era is matter dominated, the derivative of time with respect to redshift is equal to
\begin{equation}
    \frac{\de t}{\de z'} =
    \frac{1}{\sqrt{\Omega_M}H_0(1+z')^{5/2}}.
\end{equation}
The energy absorbed by the dark matter gas per unit time is equal to
\begin{equation}
    \frac{\de E_\chi}{\de t} =
    n_H \int_{0}^\infty
    \left[ \sigma^+_{\chi N}(v)\bar{E}_+(v) + 3\sigma^-_{\chi N}(v)\bar{E}_-(v)
    \right]
    v f_{\tilde{T}}(v)\; \de v,
\end{equation}
where $\bar{E}_- (v)$ and $\bar{E}_+ (v)$ are the mean energies via excitation or deexcitation interactions, respectively. These are approximated in the following way,
\begin{equation}\label{eq:energy_transfers}
\begin{split}
    \bar{E}_- (v) & = \frac{\mu v^2}{c^2}, \\
    \bar{E}_+ (v) & = \frac{\mu v^2}{c^2} + \frac{\mu}{m_\chi}E_\star.
\end{split}
\end{equation}
The dominant part in this heating process is the deexcitation energy $E_\star$. To first order, the above approximation of $\bar{E}_\pm$ is correct, as long as the increase in kinetic energy per dark matter fermion is of similar order of magnitude or smaller than $E_\star$. Due to the coupling constant limits, which are discussed in Sec.~\ref{sec:limits}, the dark matter fermion self-interactions are stronger than dark matter-baryon interactions, such that energy absorbed from the spin gas is quickly distributed in the dark matter gas.

The kinetic temperature of the hydrogen gas $T_K$ is not significantly affected by dark matter interactions for most of the considered parameter space of dark sector masses. We estimate the cooling of the hydrogen gas in the following way. As discussed above for the case of the dark matter gas, the energy transfer due to momentum exchange in an elastic dark matter-hydrogen collision is of the order $\mu v^2/c^2$. The change in the hydrogen energy due to a hyperfine excitation (or dexcitation) is of the order
\begin{equation}
    \frac{\mu}{m_H}E_\star,
\end{equation}
which is a negligible energy compared to that of the collisional momentum exchange, and can thus be ignored. The mean energy loss of hydrogen in a hydrogen-dark matter collision can thus be approximated as
\begin{equation}
    \bar{E}_H = -\frac{\mu v^2}{c^2}.
\end{equation}
The above assumes that the dark matter gas is inert, an approximation which breaks down when the dark matter gas is heated. When the dark matter velocity dispersion becomes comparable or larger than that of the hydrogen gas, cooling of the hydrogen gas is slowed down somewhat, as some collisions will actually transfer energy to the hydrogen gas. Hence the cooling of the hydrogen gas is somewhat over-estimated. In Sec.~\ref{sec:results}, we demonstrate for what dark sector masses that hydrogen gas cooling is significant. We motivate why cooling of the hydrogen gas has no effect on the spin temperature, even if the hydrogen gas is cooled by several Kelvin.

\section{Dark sector limits}\label{sec:limits}

The limits on sub-GeV dark matter is discussed thoroughly by Green and Rajendran \cite{Green:2017ybv} and Knapen et al. \cite{2017PhRvD..96k5021K}, where they consider a model with a fermion $\chi$ and a scalar mediator $\phi$, with interaction terms of the form $g_\chi \phi \bar{\chi} \chi$. While that model has spin-independent interactions, similar limits apply to the model considered in this work. In the mass ranges of interest, the strongest limits to the dark matter-hydrogen cross section, as expressed in Eq. \eqref{eq:cross_section_full}, comes from a combination of stellar cooling limits to $g_N$ and $g_e$, and dark matter self-interaction limits to $g_\chi$.

For scalar mediator particles with masses below $\sim 10~\keV$, the cooling of horizontal branch stars, red giant stars and white dwarfs sets an approximate bound of $g_N \lesssim 10^{-12}$ for hadrophilic interactions \cite{Vogel:2013raa} and $g_e \lesssim 10^{-15}$ for leptophilic interactions \cite{2015ApJ...809..141H,Hardy:2016kme}. For very light mediator masses, fifth force searches constrain these coupling constants to $g_N \lesssim 10^{-12}\times(m_V/\eV)^3$ and $g_e \lesssim 10^{-9}\times(m_V/\eV)^3$, respectively \cite{Murata:2014nra}. Fifth force limits are dominant for $m_V \lesssim 1~\eV$ in the hadrophilic case, and $m_V \lesssim 10^{-2}~\eV$ in the leptophilic case. These limits are model dependent, and differ for a spin-dependent interaction. For the hadrophilic coupling constant limits, the dominant process for stellar cooling is the Compton process $\gamma+\text{He} \rightarrow \text{He}+\phi$, involving helium. For a spin-dependent interactions, this process can be suppressed due to destructive interference of the helium core's nucleons, whose total spin is zero. Calculating new limits for spin-dependent interactions is beyond the scope of this work; similar limits apply, although with some modifications, possibly less restrictive.

In addition to the scalar mediator limits described above, the coupling constants $g_N$ and $g_e$ are subject to even stronger constraints in a pseudo-vector case, due to coupling to anomalous currents, as well as $(\text{energy}/m_V)^2$ enhanced coupling to the mediator's longitudinal mode. Coupling to anomalous currents affect for example meson decay rates; such couplings can be suppressed through the introduction of other dark sector fields \cite{2017JHEP...08..053E}. Longitudinal mode enhancement also affects mediator production rates, giving significantly stronger bounds for stellar cooling: $g_N \lesssim 10^{-17}\times (m_V/\eV)$ and $g_e \lesssim 10^{-18}\times (m_V/\eV)$ \cite{Dror:2017ehi,Dror:2017nsg}. While the enhancement of the longitudinal mode is an infra-red effect, it is contingent on the ultra-violet completion of the model and intimately connected with how the mediator's mass is generated. To suppress this enhancement, it is necessary to introduce new physics in the relevant energy scale ($\sim \MeV$ for stellar cooling); introducing new fields is strongly constrained at such low energies. In summary, constructing a more complete model that evades these constraints is beyond the scope of this work, and most likely very challenging.

There are many other limits to light dark sector particles. Dark matter interactions with protons or electrons can cause spectral distortions in the CMB \cite{Ali-Haimoud:2015pwa}, but these limits are much less restrictive than the limits discussed above. The effective number of relativistic degrees of freedom $N_\text{eff}$ during Big Bang nucleosynthesis (BBN) and recombination constrains the introduction of low mass dark sector particles. However, for couplings smaller than $g_N \lesssim 10^{-9}$, the dark sector proposed in this work is decoupled from the Standard Model before BBN \cite{2017PhRvD..96k5021K}. Another concern is the mechanism that sets the relic abundance of the dark matter fermion $\chi$. This can be accomplished via other dark sector particles or even non-thermal production, but the exact nature of this mechanism is not the primary focus of this work.

If the total dark matter abundance is constituted by the fermion $\chi$, the coupling constant $g_\chi$ will be limited by dark matter self-interaction through the Bullet Cluster. The self-interaction cross section is
\begin{equation}
	\sigma_{\chi \chi} =
    \frac{g_\chi^4 m_\chi^2}{8\pi [(m_\chi v/c)^2 + m_V^2 ]^2}.
\end{equation}
The factor of two that appears here, with respect to equation \eqref{eq:cross_section}, is due to the reduced mass of the interacting particles.

In the case of a velocity independent self-interaction, the cross section is limited to $\sigma_{\chi\chi} \lesssim 1 \times (m_\chi/\gram)~\cm^2$ \cite{Randall:2007ph}. Because we have a velocity dependent cross section, we use the 4700 km/s merger velocity of the Bullet Cluster, giving a limit
\begin{equation}
	g_\chi^4 \lesssim
    8\pi\frac{m_\chi^2(4700~\kms)^4}{c^4}\left(\frac{m_\chi}{\gram}\right)~\cm^2 \simeq 1.8\times 10^{-13}\left(\frac{m_\chi}{\MeV}\right)^3,
\end{equation}
where the mediator mass $m_V$ has been neglected.

For a sub-dominant dark matter species, self-interaction bounds are far less restrictive. As argued in \cite{Fan:2013yva}, for a dark matter species contributing $\lesssim 30$ \% to the total dark matter density, self-interaction can in principle be arbitrarily strong, such that we can set $g_\chi \simeq 1$. If this dark matter component is dissipative, by internal Bremsstrahlung emission of a light mediator, sub-structures can form, for example a thin dark disk within the Milky Way and other galaxies \cite{Fan:2013yva,Fan:2013tia}. Current bounds to such a thin dark disk limits a self-interacting sub-component to at most a few per cent of the total dark matter abundance \cite{Schutz:2017tfp,Buch:2018qdr,2018arXiv181107911W}. In our case, the mediator is not massless, such that dissipation by self-interaction is quenched below some collisional velocity threshold. For example, given a dark sector mass ratio of $m_V/m_\chi = 10^{-7}$, Bremsstrahlung is suppressed for collisional velocities smaller than $\sim 100~\kms$, such that a thin dark disk cannot form in the Milky Way.

\section{Results}\label{sec:results}

In this section, we present how the spin temperature evolves under the influence of the dark sector and compare with coupling constant limits. We assume that the asymmetric fermion $\chi$ is a dark matter sub-component, constituting ten per cent of the total dark matter abundance. As discussed in Sec.~\ref{sec:limits}, no self-interaction bounds apply, and we set $g_\chi=1$.

\begin{figure}[tbp]
\centering
\includegraphics[width=1.\textwidth,origin=c]{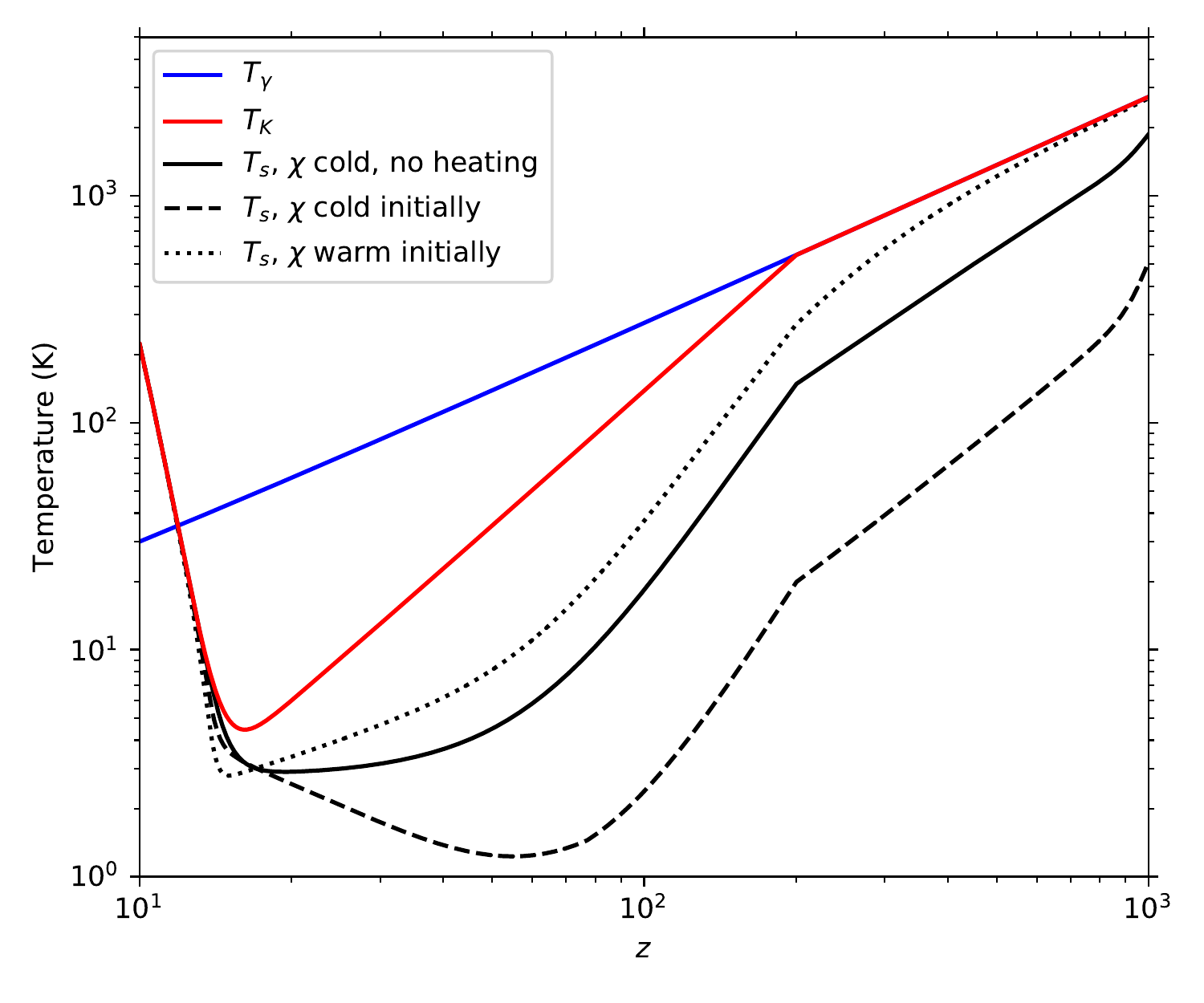}
\caption{\label{fig:Tspin2}CMB temperature ($T_\gamma$), kinetic hydrogen gas temperature ($T_K$), and spin temperature ($T_s$), as a function of redshift, where the spin temperature is affected by dark matter interactions. The spin temperature is plotted for three different cases, as described in Sec.~\ref{sec:results}. The dark matter fermion $\chi$ is assumed to constitute ten per cent of the total dark matter abundance and the masses are $m_\chi = 10~\MeV$ and $m_V = 1~\eV$. The dark sector coupling constants of the three cases are independently normalized, such that all give rise to $T_s = 3~\K$ at $z=17$.}
\end{figure}

In Fig.~\ref{fig:Tspin2} we demonstrate how the spin temperature evolves with redshift, similar to Fig.~\ref{fig:Tspin}, although under the influence of the dark sector. The spin temperature is plotted for three separate cases:
\begin{enumerate}[(a)]
    \item the dark matter gas is assumed to be cold and inert at all redshifts;
    \item the dark matter gas is inert initially, but heats with time;
    \item the dark matter gas is warm initially, such that heating is negligible.
\end{enumerate}
The mass values in this figure are taken to be $m_\chi = 10~\MeV$ and $m_V = 1~\eV$. In all three cases, the coupling constant $g_N$ is normalized such that $T_s = 3~\K$ at redshift $z = 17$, which is what the \emph{EDGES} measurement suggests. The coupling constant $g_N$ is the lowest for the ideal case (a), and the highest for the warm case (c), but differ only by a small numerical factor.

As seen in Fig.~\ref{fig:Tspin2}, the spin temperature is strongly coupled to the CMB and hydrogen gas temperatures at $z \simeq 1000$, mainly due to hydrogen-electron collisions. As the density of free electrons goes down, the spin temperature begins to be affected by the dark sector. At high redshift, cases (a) and (b) differ only due to their different coupling constant normalizations; the effective temperature is set by the gas temperature only and is proportional to $\tilde{T}\propto 1+z$, until $z\simeq 200$ where it follows $\tilde{T}\propto (1+z)^2$. For case (b), heating of the dark matter becomes significant around $z \simeq 80$, where the dark matter gas has heated enough for its velocity dispersion to become comparable to that of the hydrogen gas. For case (c), the velocity dispersion of dark matter is larger than that of hydrogen and the effective temperature is proportional to the dark matter temperature, and follows $\tilde{T}\propto (1+z)^2$ until the era of star formation.

The spin temperature evolves differently for the three cases. For case (b), where the dark matter gas is inert initially but heats up, the minimum spin temperature is found at a quite high redshift $z\simeq50$, and the spin temperature troth is very wide. For the other two cases, the minimum is located around $z\simeq17$ (depending on the details of star formation). Close to this minimum, the spin temperature is proportional to $T_s \propto 1+z$, cooling slower than the hydrogen gas temperature at this redshift. For other mass values the results are the same, with the exception of mass ratios $m_V/m_\chi \gtrsim 10^{-6}$, for which the $v^{-4}$ dependence is halted already at higher redshifts. For such high mass ratios, the minimum of the spin temperature is found at even higher redshifts and the troth is widened; this effect is especially pronounced for cases (a) and (b). The width of this troth is also dependent on the abundance of the dark matter sub-component $\chi$; a higher abundance not only permits a lower cross section, but also changes its heat capacity and makes the dark matter gas less prone to heating.

\begin{figure}[tbp]
\centering
\includegraphics[width=1.\textwidth,origin=c]{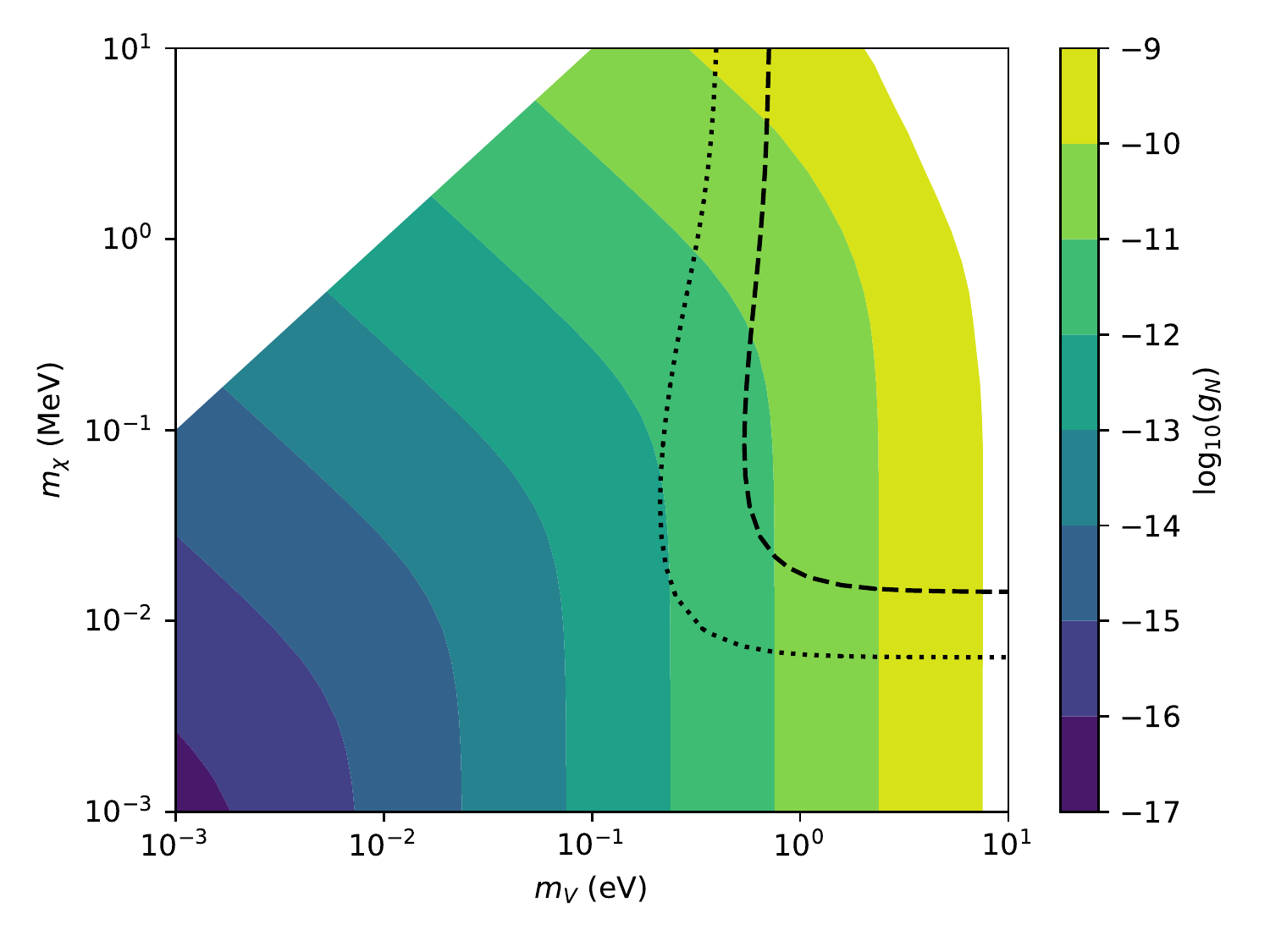}
\caption{\label{fig:gNcontour}Contour plot showing the nucleon coupling constant $g_N$ (or $g_e$, equivalently) that gives rise to a spin temperature of $T_s = 3$ K at redshift $z = 17$, for different dark matter fermion and mediator masses. In this figure, the dark matter fermion is assumed to constitute ten per cent of the total dark matter abundance, with coupling constant $g_\chi=1$. The dotted and dashed black lines correspond to where the hydrogen gas is cooled by $1~\K$ and $3~\K$, respectively.}
\end{figure}

In Fig.~\ref{fig:gNcontour}, we present the coupling constant $g_N$ or $g_e$ that gives rise to a spin temperature $T_s = 3~\K$ at redshift $z = 17$, for different dark sector masses, assuming case (b) where the dark matter gas is inert initially but heats up with time. These values can be compared to the bounds to $g_N$ and $g_e$ discussed in Sec.~\ref{sec:limits}. For low mass ratios $m_V/m_\chi$, a thin dark disk could form in the Milky Way, giving stronger limits to the dark matter fermion sub-component abundance. In this scenario, the dark matter gas would be more prone to heating due to a lower heat capacity. For this reason, mass ratios $m_V/m_\chi<10^{-8}$ are excluded in Fig.~\ref{fig:gNcontour} (upper left corner). 

In Fig.~\ref{fig:gNcontour}, the coupling constant and the mediator mass are related according to $g_N \propto m_V^2$. If coupling to anomalous currents and longitudinal mode enhancements are present, as discussed in Sec.~\ref{sec:limits}, significant reduction of the spin temperature will be excluded by many orders of magnitude in both $g_N$ and $g_e$. Even if both anomalous couplings and longitudinal mode enhancements are suppressed, other bounds are still significant. For coupling to electrons, limits are $g_e \lesssim 10^{-15}$ for mediator masses $m_V > 10^{-2}~\eV$. Given masses $m_V \simeq 10^{-2}~\eV$ and $m_\chi \simeq 1~\MeV$, the coupling constant necessary for sufficient reduction of the spin temperature is higher, but only by a relatively small numerical factor. Hence the leptophilic case seems to be excluded by a small margin, assuming that the same bounds apply for spin-dependent interactions. For coupling to nucleons, limits are $g_N \lesssim 10^{-12}$ for mediator masses $m_V > 1~\eV$. Sufficient spin temperature reduction could be achieved with a coupling constant slightly higher than $10^{-11}$.

In Fig.~\ref{fig:gNcontour}, we also show the reduction of the kinetic temperature of the hydrogen gas, as calculated in Sec.~\ref{sec:effective_temperature}. This cooling is significant for larger dark sector masses. Even for cases where the hydrogen gas is cooled by several Kelvin, this cooling does not influence the spin temperature. The spin temperature couples is a weighted mean of the CMB temperature and effective temperature of hydrogen-dark matter collisions $\tilde{T}$; the effective temperature is dominated by the heated dark matter gas, whose velocity dispersion is higher than that of the hydrogen gas.

\section{Discussion}\label{sec:discussion}

We have considered a dark matter model with spin-dependent interactions with electrons or protons, to explore if such a model could explain the 21 cm absorption signal detected by the \emph{EDGES} low-band experiment. In this model, the spin temperature of the hydrogen gas is lowered directly by spin-dependent dark matter interactions, without affecting the kinetic temperature of the hydrogen gas. This model has some merits with respect to dark matter models with spin-independent interactions. It does not need Ly$\alpha$ radiation induced coupling of the kinetic gas and spin temperatures to precede heating of the hydrogen gas. Neither is this cooling mechanism significantly affected by relative bulk motions between the hydrogen and dark matter gases. In order to evade bounds to dark matter self-interactions, it is necessary that the dark matter fermion is a sub-component constituting $\lesssim 30$ per cent of the total dark matter abundance. This is less fine-tuned than for hydrogen gas cooling by milli-charged dark matter, which is constrained to 0.3--2 per cent of the total dark matter abundance \cite{Berlin:2018sjs,Munoz:2018pzp,Munoz:2018jwq,Barkana:2018cct}.

The 21 cm absorption profile that was fitted to the \emph{EDGES} signal has the form of a well with steep walls on both sides and a flat bottom \cite{Bowman:2018yin}. The wall at lower redshift, $z\simeq15$, is due to stellar heating and Ly$\alpha$ radiation, which is dependent on details of star formation in the era before reionization and less so on the considered dark matter model. The wall at higher redshift, $z\simeq20$, is on the other hand highly dependent on the details of the dark sector. The rate of spin temperature reduction is most steep ($T_s \propto 1+z$) if the dark matter gas is slightly warm to begin with, such that the energy it absorbs from the spin gas is negligible. This model could be excluded with a precise measurement of the absorption profile, if it is indeed found to be more steep than can be accounted for with this model. For now, the exact shape of this profile is highly uncertain.

Electron or nucleon coupling with a light pseudo-vector are strongly constrained due to couplings with anomalous currents and enhanced production of the pseudo-vector's longitudinal mode, which excludes significant reduction of the spin temperature by a large margin. If anomalous couplings and longitudinal mode enhancement can be suppressed in the ultra-violet completion of the model, the model is still subject to the following limits. For dark sector coupling with electrons, it seems that significant reduction of the spin temperature is marginally excluded by bounds from stellar cooling and fifth force constraints. For dark sector coupling with protons, the case is somewhat more complicated. There are bounds to a light mediator coupling to protons, coming from cooling of horizontal branch stars and red giant stars. However, the dominant cooling process is mediator production by helium. As discussed in Sec.~\ref{sec:limits}, interactions with helium are suppressed in the case of spin-dependent interactions. For this model to give rise to a significant reduction of the spin temperature in the era before reionization, the limits to the proton coupling constant must be alleviated by one or two orders of magnitude with respect to the scalar mediator case.

In summary, we consider a novel mechanism for lowering the spin temperature before the era of cosmic reionization.  For the simple model we consider, significant spin temperature reduction is excluded by limits from stellar cooling of red giant and horizontal branch stars. Potentially, a more complete or alternative dark sector model, subject to different sets of constraints, can affect the spin temperature via the same mechanism.





\section*{Acknowledgements}
I would like to thank Garrelt Mellema, Joakim Edsj\"o, Sebastian Baum, Patrick Stengel, Sunny Vagnozzi, Jeff Dror, and Simon Knapen for useful discussions and comments.

\section*{References}
\bibliographystyle{model1-num-names}
\bibliography{refs.bib}

\begin{thebibliography}{42}
\expandafter\ifx\csname natexlab\endcsname\relax\def\natexlab#1{#1}\fi
\providecommand{\bibinfo}[2]{#2}
\ifx\xfnm\relax \def\xfnm[#1]{\unskip,\space#1}\fi
\bibitem[{Bowman et~al.(2018)Bowman, Rogers, Monsalve, Mozdzen, and
  Mahesh}]{Bowman:2018yin}
\bibinfo{author}{J.~D. Bowman}, \bibinfo{author}{A.~E.~E. Rogers},
  \bibinfo{author}{R.~A. Monsalve}, \bibinfo{author}{T.~J. Mozdzen},
  \bibinfo{author}{N.~Mahesh},
\newblock \bibinfo{title}{{An absorption profile centred at 78 megahertz in the
  sky-averaged spectrum}},
\newblock \bibinfo{journal}{Nature} \bibinfo{volume}{555}
  (\bibinfo{year}{2018}) \bibinfo{pages}{67--70}.
\bibitem[{{Hills} et~al.(2018){Hills}, {Kulkarni}, {Meerburg}, and
  {Puchwein}}]{2018arXiv180501421H}
\bibinfo{author}{R.~{Hills}}, \bibinfo{author}{G.~{Kulkarni}},
  \bibinfo{author}{P.~D. {Meerburg}}, \bibinfo{author}{E.~{Puchwein}},
\newblock \bibinfo{title}{{Concerns about Modelling of Foregrounds and the
  21-cm Signal in EDGES data}},
\newblock \bibinfo{journal}{arXiv e-prints}  (\bibinfo{year}{2018})
  \bibinfo{pages}{arXiv:1805.01421}.
\bibitem[{Feng and Holder(2018)}]{Feng:2018rje}
\bibinfo{author}{C.~Feng}, \bibinfo{author}{G.~Holder},
\newblock \bibinfo{title}{{Enhanced global signal of neutral hydrogen due to
  excess radiation at cosmic dawn}},
\newblock \bibinfo{journal}{Astrophys. J.} \bibinfo{volume}{858}
  (\bibinfo{year}{2018}) \bibinfo{pages}{L17}.
\bibitem[{Ewall-Wice et~al.(2018)Ewall-Wice, Chang, Lazio, Dore, Seiffert, and
  Monsalve}]{Ewall-Wice:2018bzf}
\bibinfo{author}{A.~Ewall-Wice}, \bibinfo{author}{T.~C. Chang},
  \bibinfo{author}{J.~Lazio}, \bibinfo{author}{O.~Dore},
  \bibinfo{author}{M.~Seiffert}, \bibinfo{author}{R.~A. Monsalve},
\newblock \bibinfo{title}{{Modeling the Radio Background from the First Black
  Holes at Cosmic Dawn: Implications for the 21 cm Absorption Amplitude}},
\newblock \bibinfo{journal}{Astrophys. J.} \bibinfo{volume}{868}
  (\bibinfo{year}{2018}) \bibinfo{pages}{63}.
\bibitem[{Fraser et~al.(2018)}]{Fraser:2018acy}
\bibinfo{author}{S.~Fraser}, et~al.,
\newblock \bibinfo{title}{{The EDGES 21 cm Anomaly and Properties of Dark
  Matter}},
\newblock \bibinfo{journal}{Phys. Lett.} \bibinfo{volume}{B785}
  (\bibinfo{year}{2018}) \bibinfo{pages}{159--164}.
\bibitem[{Pospelov et~al.(2018)Pospelov, Pradler, Ruderman, and
  Urbano}]{Pospelov:2018kdh}
\bibinfo{author}{M.~Pospelov}, \bibinfo{author}{J.~Pradler},
  \bibinfo{author}{J.~T. Ruderman}, \bibinfo{author}{A.~Urbano},
\newblock \bibinfo{title}{{Room for New Physics in the Rayleigh-Jeans Tail of
  the Cosmic Microwave Background}},
\newblock \bibinfo{journal}{Phys. Rev. Lett.} \bibinfo{volume}{121}
  (\bibinfo{year}{2018}) \bibinfo{pages}{031103}.
\bibitem[{Chluba(2015)}]{Chluba:2015hma}
\bibinfo{author}{J.~Chluba},
\newblock \bibinfo{title}{{Green's function of the cosmological thermalization
  problem – II. Effect of photon injection and constraints}},
\newblock \bibinfo{journal}{Mon. Not. Roy. Astron. Soc.} \bibinfo{volume}{454}
  (\bibinfo{year}{2015}) \bibinfo{pages}{4182--4196}.
\bibitem[{{Berlin} et~al.(2018){Berlin}, {Hooper}, {Krnjaic}, and
  {McDermott}}]{Berlin:2018sjs}
\bibinfo{author}{A.~{Berlin}}, \bibinfo{author}{D.~{Hooper}},
  \bibinfo{author}{G.~{Krnjaic}}, \bibinfo{author}{S.~D. {McDermott}},
\newblock \bibinfo{title}{{Severely Constraining Dark-Matter Interpretations of
  the 21-cm Anomaly}},
\newblock \bibinfo{journal}{Phys. Rev. Lett.} \bibinfo{volume}{121}
  (\bibinfo{year}{2018}) \bibinfo{pages}{011102}.
\bibitem[{Mu\~noz and Loeb(2018)}]{Munoz:2018pzp}
\bibinfo{author}{J.~B. Mu\~noz}, \bibinfo{author}{A.~Loeb},
\newblock \bibinfo{title}{{A small amount of mini-charged dark matter could
  cool the baryons in the early Universe}},
\newblock \bibinfo{journal}{Nature} \bibinfo{volume}{557}
  (\bibinfo{year}{2018}) \bibinfo{pages}{684}.
\bibitem[{Mu\~noz et~al.(2018)Mu\~noz, Dvorkin, and Loeb}]{Munoz:2018jwq}
\bibinfo{author}{J.~B. Mu\~noz}, \bibinfo{author}{C.~Dvorkin},
  \bibinfo{author}{A.~Loeb},
\newblock \bibinfo{title}{{21-cm Fluctuations from Charged Dark Matter}},
\newblock \bibinfo{journal}{Phys. Rev. Lett.} \bibinfo{volume}{121}
  (\bibinfo{year}{2018}) \bibinfo{pages}{121301}.
\bibitem[{Barkana et~al.(2018)Barkana, Outmezguine, Redigolo, and
  Volansky}]{Barkana:2018cct}
\bibinfo{author}{R.~Barkana}, \bibinfo{author}{N.~J. Outmezguine},
  \bibinfo{author}{D.~Redigolo}, \bibinfo{author}{T.~Volansky},
\newblock \bibinfo{title}{{Strong constraints on light dark matter
  interpretation of the EDGES signal}},
\newblock \bibinfo{journal}{Phys. Rev.} \bibinfo{volume}{D98}
  (\bibinfo{year}{2018}) \bibinfo{pages}{103005}.
\bibitem[{Pritchard and Loeb(2012)}]{Pritchard:2011xb}
\bibinfo{author}{J.~R. Pritchard}, \bibinfo{author}{A.~Loeb},
\newblock \bibinfo{title}{{21-cm cosmology}},
\newblock \bibinfo{journal}{Rept. Prog. Phys.} \bibinfo{volume}{75}
  (\bibinfo{year}{2012}) \bibinfo{pages}{086901}.
\bibitem[{{Field}(1958)}]{1958PIRE...46..240F}
\bibinfo{author}{G.~B. {Field}},
\newblock \bibinfo{title}{{Excitation of the Hydrogen 21-CM Line}},
\newblock \bibinfo{journal}{Proceedings of the IRE} \bibinfo{volume}{46}
  (\bibinfo{year}{1958}) \bibinfo{pages}{240--250}.
\bibitem[{{Allison} and {Dalgarno}(1969)}]{1969ApJ...158..423A}
\bibinfo{author}{A.~C. {Allison}}, \bibinfo{author}{A.~{Dalgarno}},
\newblock \bibinfo{title}{{Spin Change in Collisions of Hydrogen Atoms}},
\newblock \bibinfo{journal}{ApJ} \bibinfo{volume}{158} (\bibinfo{year}{1969})
  \bibinfo{pages}{423}.
\bibitem[{Zygelman(2005)}]{0004-637X-622-2-1356}
\bibinfo{author}{B.~Zygelman},
\newblock \bibinfo{title}{Hyperfine level-changing collisions of hydrogen atoms
  and tomography of the dark age universe},
\newblock \bibinfo{journal}{ApJ} \bibinfo{volume}{622} (\bibinfo{year}{2005})
  \bibinfo{pages}{1356}.
\bibitem[{{Furlanetto} and {Furlanetto}(2007)}]{2007MNRAS.374..547F}
\bibinfo{author}{S.~R. {Furlanetto}}, \bibinfo{author}{M.~R. {Furlanetto}},
\newblock \bibinfo{title}{{Spin-exchange rates in electron-hydrogen
  collisions}},
\newblock \bibinfo{journal}{MNRAS} \bibinfo{volume}{374} (\bibinfo{year}{2007})
  \bibinfo{pages}{547--555}.
\bibitem[{Furlanetto and
  Furlanetto(2007)}]{doi:10.1111/j.1365-2966.2007.11921.x}
\bibinfo{author}{S.~R. Furlanetto}, \bibinfo{author}{M.~R. Furlanetto},
\newblock \bibinfo{title}{Spin exchange rates in proton hydrogen-collisions},
\newblock \bibinfo{journal}{MNRAS} \bibinfo{volume}{379} (\bibinfo{year}{2007})
  \bibinfo{pages}{130--134}.
\bibitem[{Kuhlen et~al.(2006)Kuhlen, Madau, and Montgomery}]{Kuhlen:2005cm}
\bibinfo{author}{M.~Kuhlen}, \bibinfo{author}{P.~Madau},
  \bibinfo{author}{R.~Montgomery},
\newblock \bibinfo{title}{{The spin temperature and 21cm brightness of the
  intergalactic medium in the pre-reionization era}},
\newblock \bibinfo{journal}{ApJ} \bibinfo{volume}{637} (\bibinfo{year}{2006})
  \bibinfo{pages}{L1--L4}.
\bibitem[{Liszt(2001)}]{Liszt:2001kh}
\bibinfo{author}{H.~Liszt},
\newblock \bibinfo{title}{{The spin temperature of warm interstellar h I}},
\newblock \bibinfo{journal}{Astron. Astrophys.} \bibinfo{volume}{371}
  (\bibinfo{year}{2001}) \bibinfo{pages}{698}.
\bibitem[{Wong et~al.(2008)Wong, Moss, and Scott}]{Wong:2007ym}
\bibinfo{author}{W.~Y. Wong}, \bibinfo{author}{A.~Moss},
  \bibinfo{author}{D.~Scott},
\newblock \bibinfo{title}{{How well do we understand cosmological
  recombination?}},
\newblock \bibinfo{journal}{Mon. Not. Roy. Astron. Soc.} \bibinfo{volume}{386}
  (\bibinfo{year}{2008}) \bibinfo{pages}{1023--1028}.
\bibitem[{{Wouthuysen}(1952)}]{1952AJ.....57R..31W}
\bibinfo{author}{S.~A. {Wouthuysen}},
\newblock \bibinfo{title}{{On the excitation mechanism of the 21-cm
  (radio-frequency) interstellar hydrogen emission line.}},
\newblock \bibinfo{journal}{AJ} \bibinfo{volume}{57} (\bibinfo{year}{1952})
  \bibinfo{pages}{31--32}.
\bibitem[{{Cohen} et~al.(2017){Cohen}, {Fialkov}, {Barkana}, and
  {Lotem}}]{2017MNRAS.472.1915C}
\bibinfo{author}{A.~{Cohen}}, \bibinfo{author}{A.~{Fialkov}},
  \bibinfo{author}{R.~{Barkana}}, \bibinfo{author}{M.~{Lotem}},
\newblock \bibinfo{title}{{Charting the parameter space of the global 21-cm
  signal}},
\newblock \bibinfo{journal}{MNRAS} \bibinfo{volume}{472} (\bibinfo{year}{2017})
  \bibinfo{pages}{1915--1931}.
\bibitem[{Petraki and Volkas(2013)}]{Petraki:2013wwa}
\bibinfo{author}{K.~Petraki}, \bibinfo{author}{R.~R. Volkas},
\newblock \bibinfo{title}{{Review of asymmetric dark matter}},
\newblock \bibinfo{journal}{Int. J. Mod. Phys.} \bibinfo{volume}{A28}
  (\bibinfo{year}{2013}) \bibinfo{pages}{1330028}.
\bibitem[{Zurek(2014)}]{Zurek:2013wia}
\bibinfo{author}{K.~M. Zurek},
\newblock \bibinfo{title}{{Asymmetric Dark Matter: Theories, Signatures, and
  Constraints}},
\newblock \bibinfo{journal}{Phys. Rept.} \bibinfo{volume}{537}
  (\bibinfo{year}{2014}) \bibinfo{pages}{91--121}.
\bibitem[{{Slatyer}(2010)}]{2010JCAP...02..028S}
\bibinfo{author}{T.~R. {Slatyer}},
\newblock \bibinfo{title}{{The Sommerfeld enhancement for dark matter with an
  excited state}},
\newblock \bibinfo{journal}{Journal of Cosmology and Astro-Particle Physics}
  \bibinfo{volume}{2010} (\bibinfo{year}{2010}) \bibinfo{pages}{028}.
\bibitem[{{Tseliakhovich} and {Hirata}(2010)}]{2010PhRvD..82h3520T}
\bibinfo{author}{D.~{Tseliakhovich}}, \bibinfo{author}{C.~{Hirata}},
\newblock \bibinfo{title}{{Relative velocity of dark matter and baryonic fluids
  and the formation of the first structures}},
\newblock \bibinfo{journal}{Phys. Rev. D} \bibinfo{volume}{82}
  (\bibinfo{year}{2010}) \bibinfo{pages}{083520}.
\bibitem[{Green and Rajendran(2017)}]{Green:2017ybv}
\bibinfo{author}{D.~Green}, \bibinfo{author}{S.~Rajendran},
\newblock \bibinfo{title}{{The Cosmology of Sub-MeV Dark Matter}},
\newblock \bibinfo{journal}{JHEP} \bibinfo{volume}{10} (\bibinfo{year}{2017})
  \bibinfo{pages}{013}.
\bibitem[{{Knapen} et~al.(2017){Knapen}, {Lin}, and
  {Zurek}}]{2017PhRvD..96k5021K}
\bibinfo{author}{S.~{Knapen}}, \bibinfo{author}{T.~{Lin}},
  \bibinfo{author}{K.~M. {Zurek}},
\newblock \bibinfo{title}{{Light dark matter: Models and constraints}},
\newblock \bibinfo{journal}{Phys. Rev. D} \bibinfo{volume}{96}
  (\bibinfo{year}{2017}) \bibinfo{pages}{115021}.
\bibitem[{Vogel and Redondo(2014)}]{Vogel:2013raa}
\bibinfo{author}{H.~Vogel}, \bibinfo{author}{J.~Redondo},
\newblock \bibinfo{title}{{Dark Radiation constraints on minicharged particles
  in models with a hidden photon}},
\newblock \bibinfo{journal}{JCAP} \bibinfo{volume}{1402} (\bibinfo{year}{2014})
  \bibinfo{pages}{029}.
\bibitem[{{Hansen} et~al.(2015){Hansen}, {Richer}, {Kalirai}, {Goldsbury},
  {Frewen}, and {Heyl}}]{2015ApJ...809..141H}
\bibinfo{author}{B.~M.~S. {Hansen}}, \bibinfo{author}{H.~{Richer}},
  \bibinfo{author}{J.~{Kalirai}}, \bibinfo{author}{R.~{Goldsbury}},
  \bibinfo{author}{S.~{Frewen}}, \bibinfo{author}{J.~{Heyl}},
\newblock \bibinfo{title}{{Constraining Neutrino Cooling Using the Hot White
  Dwarf Luminosity Function in the Globular Cluster 47 Tucanae}},
\newblock \bibinfo{journal}{ApJ} \bibinfo{volume}{809} (\bibinfo{year}{2015})
  \bibinfo{pages}{141}.
\bibitem[{Hardy and Lasenby(2017)}]{Hardy:2016kme}
\bibinfo{author}{E.~Hardy}, \bibinfo{author}{R.~Lasenby},
\newblock \bibinfo{title}{{Stellar cooling bounds on new light particles:
  plasma mixing effects}},
\newblock \bibinfo{journal}{JHEP} \bibinfo{volume}{02} (\bibinfo{year}{2017})
  \bibinfo{pages}{033}.
\bibitem[{Murata and Tanaka(2015)}]{Murata:2014nra}
\bibinfo{author}{J.~Murata}, \bibinfo{author}{S.~Tanaka},
\newblock \bibinfo{title}{{A review of short-range gravity experiments in the
  LHC era}},
\newblock \bibinfo{journal}{Class. Quant. Grav.} \bibinfo{volume}{32}
  (\bibinfo{year}{2015}) \bibinfo{pages}{033001}.
\bibitem[{{Ellis} et~al.(2017){Ellis}, {Fairbairn}, and
  {Tunney}}]{2017JHEP...08..053E}
\bibinfo{author}{J.~{Ellis}}, \bibinfo{author}{M.~{Fairbairn}},
  \bibinfo{author}{P.~{Tunney}},
\newblock \bibinfo{title}{{Anomaly-free dark matter models are not so simple}},
\newblock \bibinfo{journal}{Journal of High Energy Physics}
  \bibinfo{volume}{2017} (\bibinfo{year}{2017}) \bibinfo{pages}{53}.
\bibitem[{Dror et~al.(2017{\natexlab{a}})Dror, Lasenby, and
  Pospelov}]{Dror:2017ehi}
\bibinfo{author}{J.~A. Dror}, \bibinfo{author}{R.~Lasenby},
  \bibinfo{author}{M.~Pospelov},
\newblock \bibinfo{title}{{New constraints on light vectors coupled to
  anomalous currents}},
\newblock \bibinfo{journal}{Phys. Rev. Lett.} \bibinfo{volume}{119}
  (\bibinfo{year}{2017}{\natexlab{a}}) \bibinfo{pages}{141803}.
\bibitem[{Dror et~al.(2017{\natexlab{b}})Dror, Lasenby, and
  Pospelov}]{Dror:2017nsg}
\bibinfo{author}{J.~A. Dror}, \bibinfo{author}{R.~Lasenby},
  \bibinfo{author}{M.~Pospelov},
\newblock \bibinfo{title}{{Dark forces coupled to nonconserved currents}},
\newblock \bibinfo{journal}{Phys. Rev.} \bibinfo{volume}{D96}
  (\bibinfo{year}{2017}{\natexlab{b}}) \bibinfo{pages}{075036}.
\bibitem[{Ali-Haïmoud et~al.(2015)Ali-Haïmoud, Chluba, and
  Kamionkowski}]{Ali-Haimoud:2015pwa}
\bibinfo{author}{Y.~Ali-Haïmoud}, \bibinfo{author}{J.~Chluba},
  \bibinfo{author}{M.~Kamionkowski},
\newblock \bibinfo{title}{{Constraints on Dark Matter Interactions with
  Standard Model Particles from Cosmic Microwave Background Spectral
  Distortions}},
\newblock \bibinfo{journal}{Phys. Rev. Lett.} \bibinfo{volume}{115}
  (\bibinfo{year}{2015}) \bibinfo{pages}{071304}.
\bibitem[{Randall et~al.(2008)Randall, Markevitch, Clowe, Gonzalez, and
  Bradac}]{Randall:2007ph}
\bibinfo{author}{S.~W. Randall}, \bibinfo{author}{M.~Markevitch},
  \bibinfo{author}{D.~Clowe}, \bibinfo{author}{A.~H. Gonzalez},
  \bibinfo{author}{M.~Bradac},
\newblock \bibinfo{title}{{Constraints on the Self-Interaction Cross-Section of
  Dark Matter from Numerical Simulations of the Merging Galaxy Cluster 1E
  0657-56}},
\newblock \bibinfo{journal}{ApJ} \bibinfo{volume}{679} (\bibinfo{year}{2008})
  \bibinfo{pages}{1173--1180}.
\bibitem[{Fan et~al.(2013{\natexlab{a}})Fan, Katz, Randall, and
  Reece}]{Fan:2013yva}
\bibinfo{author}{J.~Fan}, \bibinfo{author}{A.~Katz},
  \bibinfo{author}{L.~Randall}, \bibinfo{author}{M.~Reece},
\newblock \bibinfo{title}{{Double-Disk Dark Matter}},
\newblock \bibinfo{journal}{Phys. Dark Univ.} \bibinfo{volume}{2}
  (\bibinfo{year}{2013}{\natexlab{a}}) \bibinfo{pages}{139--156}.
\bibitem[{Fan et~al.(2013{\natexlab{b}})Fan, Katz, Randall, and
  Reece}]{Fan:2013tia}
\bibinfo{author}{J.~Fan}, \bibinfo{author}{A.~Katz},
  \bibinfo{author}{L.~Randall}, \bibinfo{author}{M.~Reece},
\newblock \bibinfo{title}{{Dark-Disk Universe}},
\newblock \bibinfo{journal}{Phys. Rev. Lett.} \bibinfo{volume}{110}
  (\bibinfo{year}{2013}{\natexlab{b}}) \bibinfo{pages}{211302}.
\bibitem[{Schutz et~al.(2018)Schutz, Lin, Safdi, and Wu}]{Schutz:2017tfp}
\bibinfo{author}{K.~Schutz}, \bibinfo{author}{T.~Lin}, \bibinfo{author}{B.~R.
  Safdi}, \bibinfo{author}{C.-L. Wu},
\newblock \bibinfo{title}{{Constraining a Thin Dark Matter Disk with Gaia}},
\newblock \bibinfo{journal}{Phys. Rev. Lett.} \bibinfo{volume}{121}
  (\bibinfo{year}{2018}) \bibinfo{pages}{081101}.
\bibitem[{Buch et~al.(2018)Buch, Leung, and Fan}]{Buch:2018qdr}
\bibinfo{author}{J.~Buch}, \bibinfo{author}{J.~S.~C. Leung},
  \bibinfo{author}{J.~Fan},
\newblock \bibinfo{title}{{Using Gaia DR2 to Constrain Local Dark Matter
  Density and Thin Dark Disk}},
\newblock \bibinfo{journal}{preprint}  (\bibinfo{year}{2018}).
\bibitem[{{Widmark, A.}(2019)}]{2018arXiv181107911W}
\bibinfo{author}{{Widmark, A.}},
\newblock \bibinfo{title}{{Measuring the local matter density using Gaia DR2}},
\newblock \bibinfo{journal}{A\&A} \bibinfo{volume}{623} (\bibinfo{year}{2019})
  \bibinfo{pages}{A30}.

\end{thebibliography}







\end{document}